\documentclass{article}
\usepackage[utf8]{inputenc}
\usepackage{amsmath, amssymb, amsthm, bbm, chngcntr, mathtools}
\usepackage{physics}
\usepackage{comment}
\usepackage{enumerate}
\usepackage{accents}
\usepackage{graphicx}
\usepackage[affil-it]{authblk}
\usepackage{caption}

\usepackage{subcaption}
\graphicspath{{./images/}}
\usepackage{relsize}
\usepackage{hyperref}
\usepackage{cleveref}
\hypersetup{
    colorlinks=true,
    linkcolor=blue,
    filecolor=blue,      
    urlcolor=blue,
    citecolor=blue
    }

\usepackage{booktabs, siunitx}
\usepackage[svgnames,table]{xcolor}
\usepackage[tableposition=above]{caption}
\usepackage{algorithm}
\usepackage{algpseudocode}

\usepackage{tikz-cd}
\usepackage{tikz}
\usetikzlibrary{matrix, shapes.geometric, arrows}

\tikzstyle{process} = [rectangle, rounded corners, minimum width=2.5cm, minimum height=1cm, text centered, draw=black, fill=orange!30]
\tikzstyle{arrow} = [thick,->,>=stealth]

\usepackage{fancyhdr}
\usepackage{lipsum}

\newtheorem{theorem}{Theorem}
\newtheorem{proposition}{Proposition}
\theoremstyle{definition}
\newtheorem{definition}{Definition}[section]

\title{\bfseries\LARGE Geometric Phase of Stochastic Oscillators}
\author{Yangyang Du\thanks{Department of Mathematics, University of Michigan, Ann Arbor, Michigan, USA}}
\date{\today}

\begin{document}
\maketitle

\textbf{Abstract.} Several definitions of phase have been proposed for stochastic oscillators, among which the mean-return-time (MRT) phase \cite{Schwabedal2013} and the stochastic asymptotic phase \cite{PhysRevLett.113.254101} have drawn particular attention. Quantitative comparisons between these two definitions have been done in previous studies, but physical interpretations of such a relation are still missing. In this work, we illustrate this relation using the geometric phase, which is an essential concept in both classical and quantum mechanics. We use properties of probability currents and the generalized Doob's h-transform to explain how the geometric phase arises in stochastic oscillators. Such an analogy is also reminiscent of the noise-induced phase shift in oscillatory systems with deterministic perturbation, allowing us to compare the phase responses in deterministic and stochastic oscillators. Based on our results, we hypothesize that just as in deterministic oscillators, the coupling effect among groups of stochastic oscillators has the form of a geometric phase. Furthermore, such effect can be expressed using the stochastic path integral representation of moment generating functions, which would be an interesting topic for future research.

\section{Introduction}
From action potentials to circadian rhythms, stochastic oscillators appear in various physiological models that are fundamental to computational neuroscience. Phase reduction is a popular approach in the study of oscillatory behaviors. However, the precise definition of phase can be ambivalent in the presence of noise and the properties that hold for deterministic systems can be dubious. We adapt the concept of geometric phase from physics literature to investigate the existing notions of stochastic phase, and discuss how the coupling effect may be modeled in large populations of stochastic oscillators.

An adiabatic perturbation has been shown to introduce a geometric phase-like effect compared to the unperturbed system in mesoscopic stochastic systems, which plays an important role in biological processes such as the Michaelis-Menten model for the enzymatic mechanism \cite{Sinitsyn2007}, \cite{Sinitsyn2009}. Specifically, Sinitsyn and Nemenman illustrate that the pump fluxes or ratchet currents arise from a geometric phase contribution to the moment generating functional of the fluxes. The geometric phase there is determined by the contour traced in the parameter space of the system over time, analogous to the Berry phase in quantum mechanics. More generalized definitions of the geometric phase were also developed in physics, such as the Pancharatnam phase and the Aharonov-Anandan phase, the latter of which is defined on the projective Hilbert space. We first review some results on traditional definitions of the asymptotic phase in deterministic systems. Then, we discuss the stochastic asymptotic phase and the mean-return-time phase, the relation between which has been quantitatively described in previous work \cite{Perez-Cervera2022}.

The generalized Doob's h-transform can be applied to bridge Hamiltonians and infinitesimal generators of Markovian diffusion processes, and such a connection has been discussed in the physics literature. In this work, we present a generalized Doob's h-transform applied to the backward Kolmogorov operator and investigate its effect on the phase of stochastic oscillators. This transformation also has intrinsic connection to the fluctuation-dissipation theorem. In \cite{Perez-Cervera2023}, Pérez-Cervera et al. suggests that the fluctuation-dissipation theorem can be generalized to stochastic oscillatory systems and formulated in terms of eigenfunctions of the backward Kolmogorov operator. We further elaborate this generalization using the generalized Doob's h-transform, which is an application of Girsanov's theorem.

The physical perspectives we provide here show that these notions of stochastic phases are indeed self-contained and reflect the inherent structure and property of the stochastic oscillatory systems. While \cite{Perez-Cervera2023} proposes a method for studying two weakly coupled stochastic oscillators from the perspective of a stochastic asymptotic phase, the behavior of larger coupled populations remains to be explored. Our work on the geometric contribution to the stochastic phases will allow a more systematic approach to study the coupling of stochastically perturbed oscillators.

\section{Asymptotic phase of deterministic oscillators}
The notion of an asymptotic phase is well developed for deterministic oscillatory systems, based on which phase reduction techniques have been proposed and widely used to solve computational neuroscience problems \cite{Winfree1974}\cite{winfree2001}. The phase of points that lie on a limit cycle can be calculated by a natural parametrization. For a point that is off the limit cycle but in the basin of attraction of it, an asymptotic phase can be defined, since the point gets infinitesimally close to a unique point on the limit cycle as time goes to infinity. Perturbation of the asymptotic phase by spatial noise has been studied using the phase response function, also called the infinitesimal phase response curve, or the adjoint of the limit cycle, which suggests its mathematical property. A lot of work on the phase of weakly connected oscillators is based on the following theorem proposed by Malkin \cite{Malkin1959}.

Below is a general statement of Malkin's theorem given by Hoppensteadt and Izhikevich \cite{Hoppensteadt1997}, summarized from the work of Blechman \cite{Blechman1971}, Ermentrout \cite{Ermentrout1981}, and Ermentrout and Kopell \cite{Ermentrout1991}.
\begin{theorem}
Consider a generating system
\begin{equation}
    \frac{dX}{dt} = F(X,t),\quad X\in\mathbb{R}^d,
\end{equation}
and assume that it has a $k$-parameter family of $T$-periodic solutions
\begin{equation}
    X(t) = U(t,\alpha),
\end{equation}
where $\alpha = (\alpha_1,\alpha_2,\dots,\alpha_k)^{\intercal}\in\mathbb{R}^k$. Suppose the adjoint linear system
\begin{equation}
    \frac{dZ_i}{dt} = -\nabla_x F(U(t,\alpha))^{\intercal}Z_i
    \label{adjoint}
\end{equation}
has exactly $k$ independent $T$-periodic solutions $Z_1(t,\alpha), Z_2(t,\alpha),\dots, Z_k(t,\alpha)\in\mathbb{R}^d$. Let
\begin{equation}
    Z =
    \begin{bmatrix}
    \overset{|}{\underset{|}{Z_1}} & \overset{|}{\underset{|}{Z_2}} & \dots & \overset{|}{\underset{|}{Z_n}}
    \end{bmatrix}
\end{equation}
such that
\begin{equation}
    Z^{\intercal}\nabla_{\alpha} U = I_{k\times k}.
\end{equation}
Then the perturbed system
\begin{equation}
    \frac{dX}{dt} = F(X,t) + \epsilon G(X,t)
\end{equation}
has a solution of the form
\begin{equation}
    X(t) = U(t,\alpha(\epsilon t)) + \mathcal{O}(\epsilon)
\end{equation}
where
\begin{equation}
    \frac{d\alpha}{d\tau} = \frac{1}{T}\int_0^T Z(t,\alpha)^{\intercal}G(U(t,\alpha),t)dt, \quad \tau = \epsilon t.
    \label{alpha}
\end{equation}
\end{theorem}
This theorem has been adapted to study the phase shift in groups of coupled neuronal oscillators, where $G$ represents the coupling effect due to the phase difference between individual oscillators. In such a context, we can take phase as the parameter $\alpha$ that evolves on the slow time scale represented by $\tau$. It can be shown that the phase response function $Z:=\nabla_x\phi$ satisfies \cref{adjoint}. \Cref{alpha} can be interpreted as the rate at which the asymptotic phase is shifted by the noise $G$. This noise-induced phase shift, as we will discuss below, is a special form of the so-called Berry phase.

\section{The origin of geometric phase}
The geometric phase is a fundamental concept in physics describing the phase accumulated by a system undergoing cyclic evolution in a parameter space. First introduced in quantum mechanics by Berry \cite{berry1984quantal}, the concept has since been extended to classical systems, optics, condensed matter physics, and fluid mechanics. In classical Hamiltonian systems, the analogue of the Berry phase is the Hannay angle \cite{hannay1985angle}, which characterizes the shift in action-angle variables. In dynamical systems and fluid mechanics, the geometric phase is often associated with circulation and adiabatic transport, where it manifests as a net displacement after a closed-loop evolution. Mathematically, it is frequently expressed as an integral of a connection form over a closed path or as a surface integral of the associated curvature. Following the convention used in quantum physics literature \cite{chruscinski2004geometric}, we define the Berry phase as follows.

\begin{definition}
Suppose that the adiabatic evolution of a quantum system is described by the Hamiltonian $H(R)$ where $R$ an external time-dependent parameter field. Let $|\psi_n\rangle$ be the $n$-th eigenstate that evolves according to Schrödinger's equation
\begin{equation}
    H(R(t))|\psi_n(R(t))\rangle = i\hbar|\psi_n(R(t))\rangle.
\end{equation}
Consider a closed curve $C$ in $R$. The Berry connection is
    \begin{equation}
        \mathcal{A}^{(n)} = -\operatorname{Im}\langle \psi_n(R)|\frac{\partial}{\partial R}\psi_n(R) \rangle,
    \end{equation}
and the Berry phase is
    \begin{equation}
        \gamma_C^{(n)} = \oint_C \mathcal{A}^{(n)}(R)\cdot dR.
    \end{equation}
    \label{def:Berry phase}
\end{definition}
The Berry phase is geometric in the sense that it only depends on the path in the parameter space, and is independent of gauge or how fast the path is traversed. In quantum mechanics, the magnetic field is often taken as the parameter space in order to study the magnetic effect on spins. In solid-state physics, the geometric effect on electrical polarization is an example of the Berry phase \cite{Mueller2011}. The polarization is related to the Brillouin zone, which is the fundamental region in the reciprocal space (momentum space) of a crystal, defined by the periodicity of the crystal lattice. It contains all the wavevectors that describe the electronic states in the material. Just as the change in polarization is equal to the integral of the Berry curvature over the periodic Brillouin zone, the spatial perturbation in oscillatory systems that we are interested in has the form of a Berry curvature in the conjugate space characterized by phase response functions, and therefore, as shown in \cref{alpha}, the phase shift can be expressed using the phase response function, which is periodic by construction.


\section{Asymptotic phase of stochastic oscillators}
Dissipative oscillatory systems are not Hamiltonian and their evolution is not necessarily adiabatic. Nevertheless, they display phase shifts that are very similar to Berry phase \cite{Kepler1991}. The phase variable in such systems can be taken as the equal-time parametrization of the limit cycle. In this section, we discuss two definitions of phase in stochastic oscillatory systems and relate them to a broader category of geometric phase.

Suppose that a stochastic oscillator obeys the Langevin equation
\begin{equation}
    dx = f(x)dt + g(x)dW_t
    \label{SDE1}
\end{equation}
where $f(x)\in\mathbb{R}^d$, $g(x)\in\mathbb{R}^{d\times m}$, and $W_t$ is an $m$-dimensional standard Wiener process. Let $L^{\dagger}$ be the backward Kolmogorov operator, which is the generator of the Markov process \cref{SDE1}. For any $u\in\mathcal{C}^2(\mathbb{R}^d)$,
\begin{equation}
    L^{\dagger}[u(x)] = \nabla u(x)\cdot f(x) + \sum_{i,j}D_{ij}\frac{\partial^2 u}{\partial x_i\partial x_j},
\end{equation}
where $D = \frac{1}{2}gg^{\intercal}$.

The idea of describing stochastic oscillation using the mean-return-time was first proposed in \cite{Schwabedal2013}. A section $\ell_{MRT}$ satisfies the mean-return-time (MRT) property if for all points $x \in \ell_{MRT}$, the MRT from $x$ back to $\ell_{MRT}$ is constant \cite{Cao2019} \cite{Perez-Cervera2022}. The basin of attraction $D$ is assumed to be diffeomorphic to an annulus. Let $\gamma$ be the parametrization that maps $x\in D$ to $(\alpha,\beta)\in[0, 2\pi)\times[-1, 1]$. The probability density function $P(\alpha,\beta,t)$ satisfies the Fokker-Planck equation, which can also be written as a transport equation
\begin{equation}
    \frac{\partial P}{\partial t} = L[P] = -\nabla\cdot J,
\end{equation}
where $L$ is the forward Kolmogorov operator and $J(\alpha,\beta,t)$ is known as the probability current. Let $\mathbf{j}(\alpha,\beta) = (j_{\alpha},j_{\beta})$ be the stationary probability current that corresponds to the stationary probability distribution $P_0(\alpha,\beta)$. The mean period of the oscillator can be written as
\begin{equation}
    \overline{T} = \frac{1}{\int_{-1}^{1}j_{\alpha}(\alpha,\beta)d\beta}.
\end{equation}
We can solve the Kolmogorov backward equation with a Neumann boundary condition and a jump condition \cite{Cao2019}:
\begin{align}
    L^{\dagger}T(\alpha,\beta)&=-1,\\
    \mathbf{n}_{\pm}\cdot\nabla T(\alpha,\beta) &= 0,\qquad (\alpha,\beta)\in[0,2\pi)\times\{\pm1\},\\
    \lim_{\alpha\to0^+}(T(-\alpha,\beta)-T(\alpha,\beta)) &= \overline{T}
\end{align}
where $\mathbf{n}_{\pm}$ denotes the local unit normal vector for the boundary at $(\alpha,\beta)\in[0,2\pi)\times\{\pm1\}$. The level curves of $T(\gamma(x))$ define the isochrons of mean-return-time phase.
\begin{definition}
    For any $x\in D$ and a fixed constant $T_0\in\mathbb{R}$, the mean return time (MRT) phase is defined as
    \begin{equation}
        \Theta(x) = \frac{2\pi}{\overline{T}}(T_0 - T(\gamma(x)))
    \end{equation}
    where $T$ is the solution of the Kolmogorov backward equation with boundary conditions stated above.
    \label{def:MRT phase}
\end{definition}
The MRT phase accumulates as a result of the net circulation of probability in the state space. The probability current formula has a strong resemblance to the Berry phase, and this connection is rooted in the geometric structure of quantum mechanics. A more detailed discussion can be found in \cite{Parrondo1998}.

\begin{figure}[!htbp]
    \centering
    \begin{subfigure}[b]{0.85\linewidth}
        \centering
        \includegraphics[width=\linewidth]{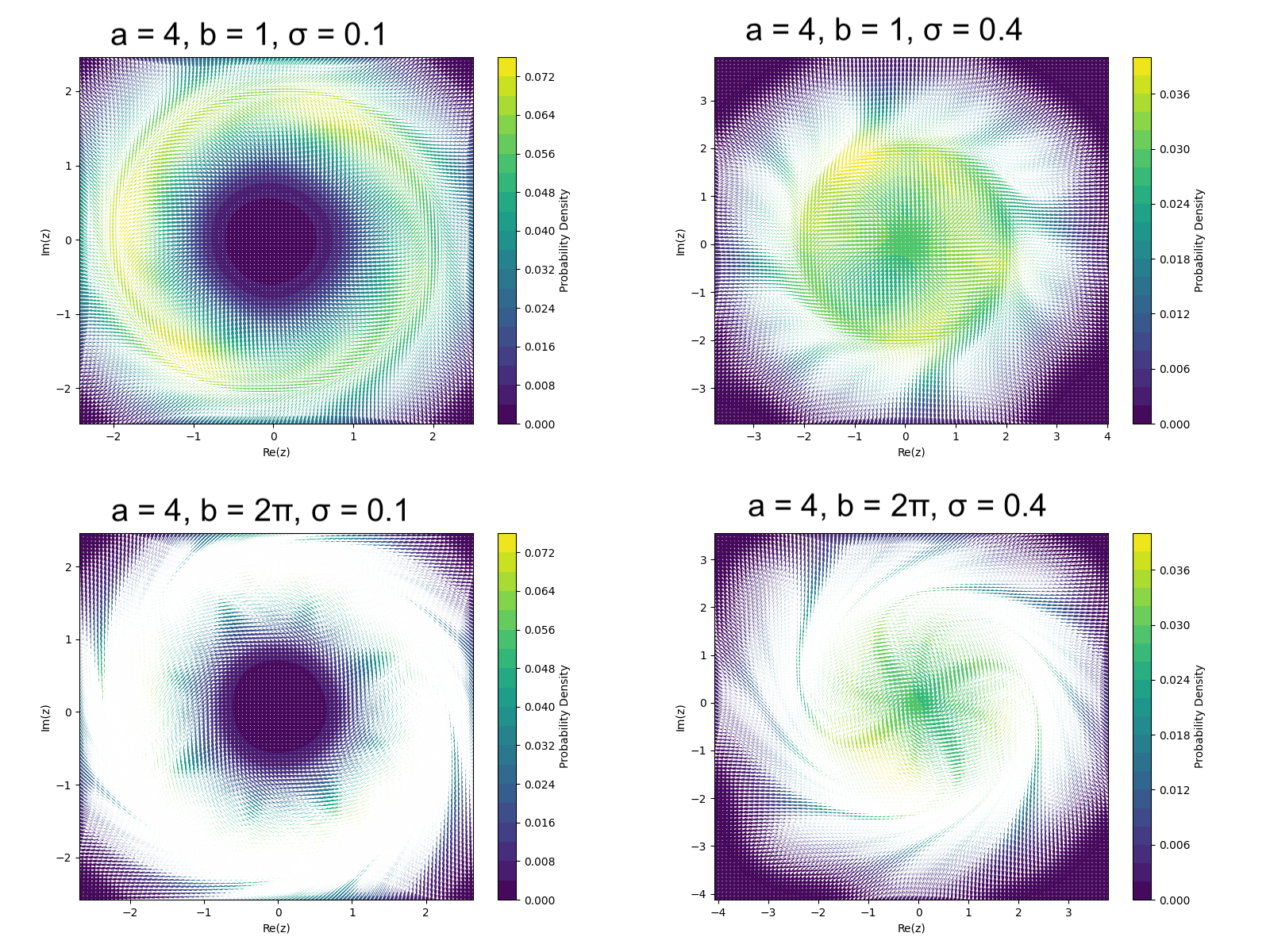}
        \caption{\textbf{Noisy Stuart-Landau oscillator, $a = 4$.}}
        \label{fig:Stuart-Landau-1}
    \end{subfigure}
    
    \begin{subfigure}[b]{0.85\linewidth}
        \centering
        \includegraphics[width=\linewidth]{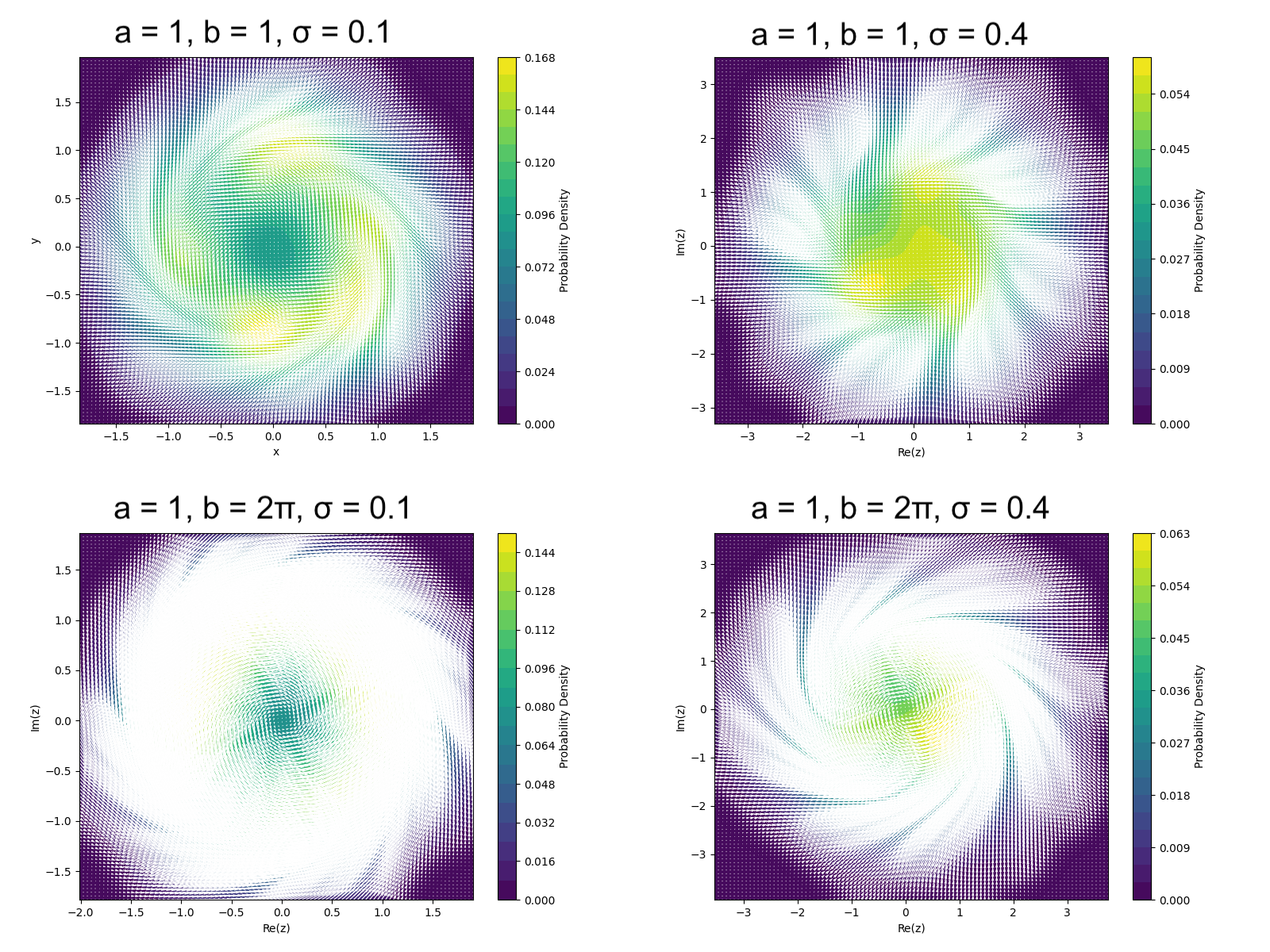}
        \caption{\textbf{Noisy Stuart-Landau oscillator, $a = 1$.}}
        \label{fig:Stuart-Landau-2}
    \end{subfigure}
    
    \caption{\textbf{Noisy Stuart-Landau oscillators.} $T=1000$, $\Delta t = 0.01$, number of samples = 1000.}
    \label{fig:Stuart-Landau}
\end{figure}

To visualize how the probability current provides an equal-time parametrization of the system, we demonstrate a Monte Carlo simulation of the noisy Stuart-Landau oscillator. The Stuart-Landau equation gives a normal form of the supercritical Hopf bifurcation. Using a complex variable $z\in \mathbb{C}$, we write the stochastic differential equation as
\begin{equation}
    \frac{dz}{dt} = (a + bi)z - |z|^2 z + \sigma \, dW_t,
\end{equation}
where $a$ is the stability parameter, $b$ is the natural frequency of the oscillator, and $\sigma$ is the noise strength. We use the Euler-Maruyama method to numerically solve the SDE, and use kernel density estimation to calculate the probability density function and the probability current. The probability current is represented as white quivers, overlaid on the contour plot that shows the probability density at the terminal time. In \cref{fig:Stuart-Landau}, we can see that this simple computation demonstrates a geometric partition of the noisy oscillatory system for different values of $a, b, \sigma$.

An alternative perspective to view the phase evolution of stochastic oscillators is through the eigenfunctions of the backward Kolmogorov operator, or the Koopman operator. In certain systems, applying spectral methods gives us a convenient and straightforward definition of stochastic phase as follows \cite{PhysRevLett.113.254101}.

\begin{definition}
    Suppose that the nontrivial eigenvalue of $L^{\dagger}$ with the least negative real part $\lambda_1 = \mu_1 + i\omega_1$ is complex (i.e., $\omega_1\neq 0$) and has algebraic multiplicity $1$. Let the polar form of the corresponding eigenfunction be $Q_{\lambda_1}^*(x) = u(x)e^{i\psi(x)}$, $u(x),\psi(x)\in\mathbb{R}$. The stochastic asymptotic phase of $x\in\mathbb{R}^d$ is $\psi(x)$.
    \label{backward phase}
\end{definition}
In \cite{Perez-Cervera2022}, a quantitative relation is derived between the MRT phase and the stochastic asymptotic phase. The expression is given as follows.
\begin{proposition} Let $\Theta(x)$ and $\psi(x)$ denote the mean-return-time phase and the stochastic asymptotic phase, respectively. Let $\overline{T}$ be the mean return time, and $\omega_1 = \arg[\lambda_1]$. Then,
\begin{align}
    L^{\dagger}[\Theta(x)] &= L^{\dagger}[\psi(x)] + \Omega(x) + \Delta\omega, \\
    \Omega(x) &:= 2\sum_{i,j}D_{ij}\frac{\partial \ln u}{\partial x_i}\frac{\partial \psi}{\partial x_j}, \label{Omega}\\
    \Delta\omega &:=\frac{2\pi}{\overline{T}} - \omega_1.
\end{align}
\label{quantative_comparison}
\end{proposition}
\Cref{Omega} follows from a direct calculation in \cite{Perez-Cervera2022} which uses the \textit{carre du champs} operator of the diffusion process. The difference in frequencies, $\Delta\omega$, is due to the different reference frames used in the two definitions. As we discussed above, the geometric contribution to $\Theta(x)$ results from the probability current. Next, we discuss the analogue of the geometric phase in the evolution of $\Psi(x)$.

Given two nonorthogonal Hilbert states $\psi_0$ and $\psi_1$, the Pancharatnam phase $\arg \langle \psi_0 | \psi_1 \rangle$ measures the relative phase between them. Originally proposed to compare different polarization states of light, the Pancharatnam phase can also be used for geometric interpretation of quantum systems, and can be viewed as a discrete counterpart of the Berry phase. The geometric phase can be obtained by subtracting the dynamical phase from the Pancharatnam phase, which gives rise to the definition of the Aharonov-Anandan phase.

Aharonov-Anandan phase \cite{aharonov1987phase} is an extension of the Berry phase in that it removes the adiabatic assumption and replaces the parameter space with the projective Hilbert space. It is particularly useful in cases where the explicit formula for curvature might be missing.

\begin{definition}
Suppose that the trajectory of a quantum state $\psi(t)\in\mathcal{H}$ is closed in the projective Hilbert space $P(\mathcal{H})$, and there exists some $T>0$ such that $\pi(\psi(0)) = \pi(\psi(T))$ where $\pi: \mathcal{H}\rightarrow P(\mathcal{H})$ is the canonical projection. The Aharonov-Anandan phase is defined by 
\begin{equation}
    \phi_{AA} = \arg \langle \psi(0) | \psi(T) \rangle - (-\frac{1}{\hbar}\int_0^T  \langle \psi(t) | H | \psi(t) \rangle dt).
\end{equation}
\end{definition}
In the above equation,
\begin{align}
    \phi_{total} &= \arg \langle \psi(0) | \psi(T) \rangle,\\
    \phi_{dyn} &= - \frac{1}{\hbar}\int_0^T  \langle \psi(t) | H | \psi(t) \rangle dt.
\end{align}
Now we want to show that the evolution of the stochastic asymptotic phase follows the same pattern. Specifically, in the following equation
\begin{equation}
    \Omega(x(t)) = \omega_1 - L^{\dagger}[\psi(x(t))],
\end{equation}
$\omega_1$, $L^{\dagger}[\psi(x(t))]$ and $\Omega(x(t))$ correspond to the total, dynamical, and geometric angular velocities, respectively. First, we can calculate the Pancharatnam phase using the autocorrelation function. It can be shown that the autocorrelation function of $Q_{\lambda}^*$ satisfies \cite{Perez-Cervera2023}
\begin{align}
    C_{\lambda,\lambda}(\tau) &= \langle Q_{\lambda}^*(x(\tau))Q_{\lambda}^*(x(0))\rangle\\
    &= \langle |Q_{\lambda}^*|^2(x(\tau))\rangle e^{\lambda\tau}.
\end{align}
Therefore,
\begin{equation}
    \arg \langle \psi(0) | \psi(\frac{2\pi}{\omega_1}) \rangle = 2\pi.
\end{equation}
In the quantum mechanical system,
\begin{align}
    \frac{d\phi_{dyn}}{dt} &= -\frac{1}{\hbar} \langle \psi | H | \psi \rangle\\
    &= \arg[\frac{1}{i\hbar}\langle \psi | H | \psi \rangle],
\end{align}
where the inner product $\langle \psi | H | \psi \rangle$ is the expectation of the Hamiltonian. Recall that the Schrödinger equation can be written as
\begin{equation}
    \frac{d}{dt}|\psi(t)\rangle = \frac{1}{i\hbar}\cdot H|\psi(t)\rangle.
\end{equation}
In our dissipative system, the evolution of the expectation of observables is governed by the backward Kolmogorov operator. Therefore, the stochastic asymptotic phase acts as the dynamical phase:
\begin{equation}
    \frac{d\phi_{dyn}}{dt} =\frac{d}{dt}\mathbb{E}[\psi(x)] = L^{\dagger}[\psi(x)].
\end{equation}

The connection between the geometric phase in the Schrödinger equation and the Kolmogorov equation is more implicit. We illustrate this equivalence from the perspective of a generalized definition of the phase response function. In the deterministic case, the phase response function describes the effect of a perturbation that pushes the vector off the limit cycle. In the stochastic system, we need to account for perturbations that drive the system away from its dominant eigenstate. Therefore, similar to the deterministic phase equation, there is an additional drift term which arises from such conditioning. We can use the generalized Doob's $h$-transform to compute the conditioned flow.

Following \cite{Chetrite2015}, we define the generalized Doob's transform as follows.
\begin{definition}
    Let $h$ be a strictly positive function on a Polish space $\mathcal{E}$ and $f$ an arbitrary function on the same space. The generalized Doob transform of the process $X_t$ with generator $L$ is the new process with generator
    \begin{equation}
        L^{h,f}\equiv h^{-1}Lh-f,
    \end{equation}
    \begin{equation}
        L^{h,f}[\phi(x)] = h^{-1}(x)\cdot L[h(x)\cdot \phi(x)] - f(x)\cdot\phi(x).
    \end{equation}
\end{definition}
In the classical Doob's transform, $h$ is usually chosen to be an eigenfunction of the operator $L$, and $f$ the constant function whose value is equal to the corresponding eigenvalue. In particular, if $Lh = 0$, then $h$ is called a harmonic function, and the transformed process is conservative. However, in the generalized Doob's transform, we can also let $f(x):=h(x)^{-1}L[h(x)]$ to make the new Markovian process conservative.

The following proposition shows how the generalized Doob's h-transforms diffusion processes.
\begin{proposition}
    Suppose a diffusion process $X_t\in\mathbb{R}^d$ follows
    \begin{equation}
        dX_t = b(X_t)dt + \sigma(X_t)dW_t.
    \end{equation}
    where $b(X_t)\in\mathbb{R}^d$, $\sigma(X_t)\in\mathbb{R}^{d\times m}$, and $W_t$ is an $m$-dimensional standard Wiener process.
    For a positive function $h(x)$, the generalized Doob's $h$-transformed process $Y_t$ follows
    \begin{equation}
        dY_t = \left( b(Y_t) + \sigma(Y_t) \sigma^T(Y_t) \nabla \ln h(Y_t) \right) dt + \sigma(Y_t) dW_t^*.
    \end{equation}
    where $W_t^*$ is also an $m$-dimensional standard Wiener process under the transformed measure.
\end{proposition}
Using $h(x) = u(x) = |Q_{\lambda}^*(x)|$ and $f = h^{-1}(Lh)$, we get the generator of the transformed process
\begin{equation}
    (L^{\dagger})^{u,f}[\psi] = L^{\dagger}[\psi] + 2\sum_{i,j}D_{ij}\frac{\partial \ln u}{\partial x_i}\frac{\partial \psi}{\partial x_j}.
\end{equation}
In other words, the transformed process is a diffusion with the same noise as the original diffusion, but with a modified drift. The additional drift due to the conditioning on $u(x)$ is exactly $\Omega(x)$.

The generalized Doob's transform applied here is akin to a gauge transformation in quantum mechanics, the idea of which can be traced back to the early establishment of potential theory. For example, Albeverio et al. \cite{Albeverio1977} proved that the Hamiltonian operator on a dense domain of $L^2(\mathbb{R}^d,d\mu)$ can be expressed as
\begin{equation}
    H = -\Delta-2\nabla\ln(\phi(x))\cdot\nabla
    \label{hamiltonian_Doob}
\end{equation}
where $\Delta$ is the Laplacian and
\begin{equation}
    d\mu(x)=\phi(x)^2dx.
\end{equation}
$\phi$ is usually taken to be the ground state wave function. The generator of the modified Markov process $L_H = -H_{\phi}$ has an additional drift term $2\nabla\ln(\phi(x))\cdot\nabla$. The ground state transformation is a unitary map:
\begin{align*}
    U_{\phi}: L^2(\mathbb{R}^d,dx) &\longrightarrow L^2(\mathbb{R}^d,d\mu)\\
    f &\longmapsto \frac{f}{\phi}
\end{align*}
that acts on the Hamiltonian by $H_{\phi} = U_{\phi}HU_{\phi}^{-1}$. Similar approaches have been used to study the $N$-body Hamiltonian, e.g., the Gross-Pitaevskii equation \cite{Albeverio2015}, which is a nonlinear Schrödinger equation that describes the dynamics of a Bose-Einstein condensate in the mean-field approximation and incorporates interactions between particles. Again, we replaced the Hamiltonian operator with the backward Kolmogorov operator in our study of stochastic oscillators. Moving forward, it would be of great interest to see if similar techniques can be applied to N-coupled stochastic oscillators with mean-field coupling.

The physical interpretation of using Doob's h-transform is that the state vector stays in the $n$-th eigenspace due to the adiabatic theorem. In our case, it corresponds to the assumption that $u(x)e^{i\psi(x)}$ remains the dominant eigenfunction of the backward Kolmogorov operator. Writing the eigenfunction in the polar form allows us to interpret $u(x)$ as a probability amplitude function (up to normalization), which is the main idea that lies behind the change of measure in Doob's h-transform. This transformation also plays a significant role in the generalized fluctuation-dissipation theorem, which describes the response of a system to small perturbations. In \cite{Perez-Cervera2023}, the authors highlight a similarity to the fluctuation-dissipation theorem based on the linear-response function. We can also see this connection using the Doob's h-transform.

The fluctuation-response theorem connects the response of a system to small perturbations with its equilibrium fluctuations. Originally constrained to Hamiltonian perturbations of equilibrium systems, it can be extended to general Markovian processes by relating action functionals to exponential martingales through the Radon-Nikodym derivative of the tilted path measure with respect to the original measure \cite{Chetrite2011}. With the generalized Doob's transform which generates the tilted Markovian process and the Radon-Nikodym derivative, the fluctuation relations can be yielded from a first-order Taylor expansion of the exponential martingale identity.

\section{Conclusion}
In a deterministically perturbed oscillatory system, the geometric phase can be expressed as the inner product of the phase response function and the perturbation. We studied different definitions of stochastic phases. The mean-return-time phase is closely related to the probability current, which has the same form of Berry phase. The stochastic asymptotic phase alone can be considered as a dynamical phase contributed by the original generator of the diffusion, i.e. the Kolmogorov backward diffusion, but the expression of its evolution requires an additional drift term, which can be derived from a generalized Doob's transform, and can be comprehended as an Aharonov-Anandan phase. In both cases, the evolution of the phase has a geometric part in the sense that it is a time-independent path integral.

By providing a qualitative explanation of the connections between these concepts, our work demonstrates the consistency of notions of phase called upon in different scenarios. Future work can be done on exploring larger systems of coupled stochastic oscillators, which may involve more complex behaviors such as synchronization. As the coupling between deterministic oscillators can be expressed as a geometric phase, we expect that probability current and the generalized Doob's h-transform can facilitate further studies of coupled stochastic oscillators and will enhance our understanding of the mechanism of large neural networks.

\bibliographystyle{plain}
\bibliography{sn-bibliography}

\begin{thebibliography}{10}

\bibitem{aharonov1987phase}
Y.~Aharonov and J.~Anandan.
\newblock Phase change during a cyclic quantum evolution.
\newblock {\em Physical Review Letters}, 58(17):1593--1596, 1987.

\bibitem{Albeverio1977}
S.~Albeverio, R.~Høegh-Krohn, and L.~Streit.
\newblock Energy forms, hamiltonians, and distorted brownian paths.
\newblock {\em Journal of Mathematical Physics}, 18(5):907--917, 1977.

\bibitem{Albeverio2015}
S.~Albeverio and S.~Ugolini.
\newblock A doob h-transform of the gross–pitaevskii hamiltonian.
\newblock {\em Journal of Statistical Physics}, 161(2):486--508, 2015.

\bibitem{berry1984quantal}
M.~V. Berry.
\newblock Quantal phase factors accompanying adiabatic changes.
\newblock {\em Proceedings of the Royal Society of London. Series A, Mathematical and Physical Sciences}, 392(1802):45--57, 1984.

\bibitem{Blechman1971}
I.~I. Blechman.
\newblock {\em Synchronization of Dynamical Systems}.
\newblock Nauka, Moscow, 1971.
\newblock [in Russian: "Sinchronizatzia Dinamicheskich Sistem"].

\bibitem{Cao2019}
A.~Cao, B.~Lindner, and P.~J. Thomas.
\newblock A partial differential equation for the mean--return-time phase of planar stochastic oscillators.
\newblock {\em SIAM Journal on Applied Dynamical Systems}, 18(3):1454--1484, 2019.

\bibitem{Chetrite2011}
R.~Chetrite and S.~Gupta.
\newblock Two refreshing views of fluctuation theorems through kinematics elements and exponential martingale.
\newblock {\em Journal of Statistical Physics}, 143(3):543--565, 2011.

\bibitem{Chetrite2015}
R.~Chetrite and H.~Touchette.
\newblock Nonequilibrium markov processes conditioned on large deviations.
\newblock {\em Annales Henri Poincaré}, 16:2005--2057, 2015.

\bibitem{chruscinski2004geometric}
D.~Chruściński and A.~Jamiołkowski.
\newblock {\em Geometric Phases in Classical and Quantum Mechanics}, volume~36 of {\em Progress in Mathematical Physics}.
\newblock Birkhäuser, 2004.

\bibitem{Ermentrout1981}
G.~B. Ermentrout.
\newblock n : m phase-locking of weakly coupled oscillators.
\newblock {\em Journal of Mathematical Biology}, 12:327--342, 1981.

\bibitem{Ermentrout1991}
G.~B. Ermentrout and N.~Kopell.
\newblock Multiple pulse interactions and averaging in systems of coupled neural oscillators.
\newblock {\em Journal of Mathematical Biology}, 29:195--217, 1991.

\bibitem{hannay1985angle}
J.~H. Hannay.
\newblock Angle variable holonomy in adiabatic excursion of an integrable hamiltonian.
\newblock {\em Journal of Physics A: Mathematical and General}, 18(2):221--230, 1985.

\bibitem{Hoppensteadt1997}
F.~C. Hoppensteadt and E.~M. Izhikevich.
\newblock {\em Weakly Connected Neural Networks}.
\newblock Springer, New York, 1997.

\bibitem{Kepler1991}
T.~B. Kepler and M.~L. Kagan.
\newblock Geometric phase shifts under adiabatic parameter changes in classical dissipative systems.
\newblock {\em Physical Review Letters}, 66(7):847--849, 1991.

\bibitem{Malkin1959}
I.~G. Malkin.
\newblock {\em Some Problems in the Theory of Nonlinear Oscillations}.
\newblock United States Atomic Energy Commission, Washington, D.C., 1959.

\bibitem{Mueller2011}
E.~Mueller.
\newblock Basic training in condensed matter physics-semiclassics, 2011.
\newblock Accessed: Feb 23, 2025.

\bibitem{Parrondo1998}
J.~M.~R. Parrondo.
\newblock Reversible ratchets as brownian particles in an adiabatically changing periodic potential.
\newblock {\em Physical Review E}, 57(6):7297--7300, 1998.

\bibitem{Perez-Cervera2023}
A.~P{\'e}rez-Cervera, V.~V. Gutkin, P.~J. Thomas, and B.~Lindner.
\newblock A universal description of stochastic oscillators.
\newblock {\em Proceedings of the National Academy of Sciences}, 120(32):e2303222120, 2023.

\bibitem{Perez-Cervera2022}
A.~P{\'e}rez-Cervera, B.~Lindner, and P.~J. Thomas.
\newblock Quantitative comparison of the mean--return-time phase and the stochastic asymptotic phase for noisy oscillators.
\newblock {\em Biological Cybernetics}, 116(2):219--234, 2022.

\bibitem{Schwabedal2013}
J.~T.~C. Schwabedal and A.~Pikovsky.
\newblock Phase description of stochastic oscillations.
\newblock {\em Physical Review Letters}, 110(20):204102, 2013.

\bibitem{Sinitsyn2009}
N.~A. Sinitsyn.
\newblock Stochastic pump effect and geometric phases in dissipative and stochastic systems.
\newblock {\em Journal of Physics A: Mathematical and Theoretical}, 42(19):193001, 2009.

\bibitem{Sinitsyn2007}
N.~A. Sinitsyn and I.~Nemenman.
\newblock Universal geometric theory of mesoscopic stochastic pumps and reversible ratchets.
\newblock {\em Physical Review Letters}, 99(22):220408, 2007.

\bibitem{PhysRevLett.113.254101}
P.~J. Thomas and B.~Lindner.
\newblock Asymptotic phase for stochastic oscillators.
\newblock {\em Phys. Rev. Lett.}, 113:254101, Dec 2014.

\bibitem{Winfree1974}
A.~T. Winfree.
\newblock Patterns of phase compromise in biological cycles.
\newblock {\em J. Math. Biol.}, 1:73--95, 1974.

\bibitem{winfree2001}
A.~T. Winfree.
\newblock {\em The Geometry of Biological Time}.
\newblock Springer, New York, 2nd edition, 2001.

\end{thebibliography}

\end{document}